\documentclass[12pt]{article}
\usepackage{epsfig, cite}
\usepackage{amsfonts}
\usepackage{amsopn}
\usepackage{amsmath}
\usepackage{verbatim,amsthm}

\title
{Generalized Hooke Law for Relativistic Membranes and p-branes}
\author{A. A. Zheltukhin 
\thanks{e-mail: aaz@physto.se}  \\  
Kharkov Institute of Physics and Technology, \\
1, Akademicheskaya St., Kharkov, 61108, Ukraine}

\date{}

\begin{document}

\maketitle

\begin{abstract}
The character of elastic forces of relativistic membranes 
and $p$-branes encoded in their nonlinear equations is studied.  
 The toroidal brane equations are reduced to the classical equations 
 of anharmonic elastic media described by monomial 
potentials. Integrability of the equations is discussed and 
some of their exact solutions are constructed.  
 
\end{abstract} 

\section{Introduction}

A.I. Akhiezer paid much attention to search for effects connected with elastic
 wave propagation in condensed matter physics \cite{ABP}. 
 Relativistic membranes (p=2) and $p$-branes in higher dimensional space-time 
 are fundamental objects of string theory \cite{M1}, and their macroscopic physics is also 
 controlled  by effective elastic forces of fluxes of elementary particle fields,
 like QCD tubes in string theory. However, quantization of branes is blocked up 
 by nonlinearity of their equations 
 (see e.g. [3-15] \nocite{tucker, hoppe1, BST, DHIS, FI, WHN, Z_0, WLN, BZ_0, hoppe2, Pol, UZ, hoppe6} 
 and refs. there). The classical and quantum problems of the brane physics deserve great attention
 and stimulate investigation of the elastic forces associated with relativistic branes.
 
 Here we search the physics of closed $p$-branes (with $p=2,3,...,(D-1)/2$)
 evolving in \mbox{$D=(2p+1)$}-dimensional Minkowski space, 
and find their exact solutions. 
The brane shape is chosen to be invariant under the 
global symmetry $O(2)\times O(2)\times...\times O(2)$.
The $p$-brane equations are reduced to nonlinear ones
of an anharmonic elastic medium with a symmetric stress tensor
generated by the interaction Hamiltonian proportional to a monomial
 potential of the degree $2p$.
Exact solvability of degenerate p-brane shaped as  p-torus
with equal radii, is established.
The found solutions are presented by (hyper)elliptic functions 
that describe p-branes contracting during the time defined by
 their energy density and the dimension $p$.

\section{p-branes as elastic media}

In the orthogonal gauge $(\dot{\vec{x}} \cdot \partial_r \vec{x})=0$, 
with the index $r$
numerating the space-like $p$-brane parameters $\sigma^{r} \  (r=1,2,...,p)$, the
equations of $p$-brane in $D$-dimensional Minkowski space 
 are transformed into the second-order PDE for its  \mbox{$(D-1)$}-dimensional 
 Euclidean vector $\vec{x}$ (see [16-18] \nocite{JU1, TZ, ZT} and refs. there)
\begin{equation}\label{xeqv}
 \ddot{\vec{x}}=\frac{T}{\mathcal{P}_0}\partial_r \left( \frac{T}{\mathcal{P}_0}|g|g^{rs}\partial_s \vec{x}\right),\, \,
 \dot{\mathcal{P}_0}=0,
\end{equation}
where \mbox{$\dot{\vec{x}}\equiv\partial_{t}\vec{x}$}, the energy 
density ${\mathcal{P}_0}=T\sqrt{\frac{|g|}{1-\dot{\vec{x}}^2}}$ with \mbox{$g=\det(g_{rs})$},
the induced metric $g_{rs}=\partial_r \vec{x} 
\cdot\partial_s \vec{x}$
on the p-brane hypersurface $\Sigma_{p}$, and the brane tension $T$.
The system (\ref{xeqv}) is rather complicated and its general solution 
is not known. Thus, to get an  information on brane physics one can 
study particular solutions of (\ref{xeqv}). To find such solutions for 
 branes with a fixed dimension $p$ ($p=2,3,...,(D-1)/2$) we fix the 
Minkowski space dimension to be odd \mbox{$D=2p+1$}. 
Moreover, we suppose that 
the closed brane hypersurface $\Sigma_{p}$ is invariant under 
the global symmetry $O(2)\times O(2)\times...\times O(2)$. 
Then, using the residual gauge symmetry of the orthonormal gauge  
\begin{equation}\label{diff}
\tilde{t}=t, \ \ \ \ \tilde{\sigma}^r=f^r(\sigma^s)
\end{equation}
we present the 
Euclidean $p$-brane vector $\vec{x}(t,\sigma^r)$ as
\begin{eqnarray}\label{gmtr}  
\vec{x}^T=(q_1\cos\sigma^1,q_1\sin\sigma^1, \ldots , q_p\cos\sigma^p,q_p\sin\sigma^p),\label{ganz} 
\end{eqnarray}
using the polar coordinate pairs $(q_{r}(t),\sigma^r)$. 
It results in the diagonalized metric $g_{rs}(t)$ independent of $\sigma^r$
\begin{eqnarray}\label{metr} 
g_{rs}(t)= q_{r}^{2}(t)\delta_{rs}, \ \
 g=(q_{1} q_{2}...q_{p})^{2}.
\end{eqnarray}
The anzats (\ref{gmtr}) shows that the coordinates $\mathbf{q}(t)=(q_{1},q_{2},...,q_{p})$ 
are the time-dependent radii $\mathbf{R}(t)=(R_{1},R_{2},\ldots,R_{p})$  
of the flat $p$-torus $\Sigma_{p}$. 
As a consequence, the energy density $ \mathcal{P}_0$  
 becomes independent of the $p$-torus parameters $\sigma^{r}$ 
 and reduces to a constant C chosen to be positive 
\begin{equation}\label{gHC} 
\mathcal{P}_0\equiv   
T\sqrt{\frac{(q_{1}q_{2} \ldots q_{p})^2}{1- \dot{\mathbf{q}}^2}} = C.
\end{equation}  
It means that the Hamiltonian density $\mathcal{H}_0$ corresponding 
to $\vec{x}$ (\ref{gmtr}) equals the constant $C$
\begin{equation}\label{gHdns} 
\mathcal{H}_0=\mathcal{P}_0=
\sqrt{\boldsymbol{\pi}^2+T^{2}(q_{1}q_{2}...q_{p})^{2}}, 
\end{equation}
 where $\boldsymbol{\pi}(t)=(\pi_{1}, \pi_{1},..., \pi_{p})$ 
 is the canonical momentum conjugate to $\mathbf{q}(t)$.
Then Eqs. (\ref{xeqv}) are reduced to the  PDE equations for the  
world vector $\vec{x}(t, \sigma^r)$
\begin{eqnarray}\label{xeqvg}
 \ddot{\vec{x}} - (\frac{T}{C})^2 gg^{rs} \partial_{rs} \vec{x}=0
\end{eqnarray}
which are equivalent to the algorithmic chain of $p$ 
nonlinear equations for the components $q_{1},q_{2},...,q_{p}$ 
\begin{equation}\label{qeqv} 
\ddot{q_{r}} + (\frac{T}{C})^2 (q_{1}\ldots 
q_{r-1} q_{r+1}\ldots q_{p})^{2} q_{r} =0, 
\end{equation}
where the component index $r$ runs from 1 to p.

The first integral of the system  (\ref{qeqv}) is given by
the relation (\ref{gHC}) presented in the form 
\begin{equation}\label{1sti}
\dot{\mathbf{q}}^2+ (\frac{T}{C})^2
(q_{1} q_{2}...q_{p})^{2} = 1.
\end{equation}
 Eqs. (\ref{qeqv}) are presented
 in a compact form as
\begin{equation}\label{veqv} 
C\ddot{\mathbf{q}}=- \frac{\partial V}{\partial\mathbf{q}}, 
\end{equation}
with the elastic energy density  $V(\mathbf{q})$ proportional 
to the determinant $g$ of the metric tensor of $\Sigma_p$
\begin{equation}\label{poten} 
V(\mathbf{q})=\frac{T^2}{2C} g \equiv
\frac{T^2}{2C}(q_{1}...q_{p})^{2}. 
\end{equation}

The equations of an elastic nonrelativistic medium with 
the mass density $\rho$ have the form  \cite{LL} 
\begin{equation}\label{lali} 
\rho\ddot{u}_{i}= \frac{\partial\sigma_{ik}}{\partial x_{k}},
\end{equation}
where $\ddot{u}_{i}$ and $\sigma_{ik}$ are the medium
acceleration and the stress tensor, respectively.
 Then, one can see that Eqs. (\ref{veqv}) are presented in the form  of Eqs. (\ref{lali})
\begin{equation}\label{eleqv} 
C\ddot{q_{r}} =-\frac{T^2}{2C}\delta_{rs}
 \frac{\partial g}{\partial q_{s}}
\end{equation}
with the symmetric stress tensor $\sigma_{rs}$ defined as 
\begin{equation}\label{strst} 
\sigma_{rs}=-p\delta_{rs}, \ \ \ \ \ \ \
p=\frac{T^2}{2C} g \equiv\frac{T^2}{2C} \prod_{s=1}^{p}q_{s}^{2}.
\end{equation}
 The relations (\ref{strst}) show that $p=V$,
  $p$ is an isotropic pressure per unit (hyper)area of the p-brane
  (hyper)volume, and the constant $C$
is a relativistic generalization of the rest mass density  $\rho$.  
The pressure $p$ is created by the elastic force $F_r$ 
\begin{equation}\label{force} 
F_{r}= - \frac{\partial V}{\partial q_{r}}\equiv-\frac{T^2}{C} 
(q_{1}\ldots q_{r-1} q_{r+1}\ldots q_{p})^{2} q_{r}. 
\end{equation}
Relation (\ref{force}) yields an anharmonic generalization of the 
 Hooke law for the toroidal $p$-brane elasticity.  

Taking into account that in the totally fixed gauge 
the Hamiltonian density (\ref{gHdns})
reduces to the  constant $C$, 
 one can introduce  a new Hamiltonian density 
 $$
\mathcal{H}=\frac{\mathcal{H}_{0}^{2}}{2C}=C/2
$$
 quadratic in the brane momentum $\boldsymbol{\pi}$.
The brane Hamiltonian $H$ associated with the density $\mathcal{H}$ is 
\begin{eqnarray}\label{nosqrham} 
H=\int d^p\sigma \mathcal{H}, \ \ \ 
\mathcal{H}=\frac{1}{2C} (\boldsymbol{\pi}^2 + T^{2} (q_{1}...q_{p})^{2}).
\end{eqnarray}
As a result, Eqs. (\ref{qeqv}) are  presented in the Hamiltonian form 
with the standard PBs 
$$
 \{\pi_{a}, q_b\}=\delta_{ab}, \ \ \  \{q_{a}, q_b \}=0, \ \ \
 \{\pi_{a}, \pi{_b} \}=0. 
$$
 The family of the Hamiltonians (\ref{nosqrham}) contains
  the potential 
 energy terms quartic in $\mathbf{q}$ for membranes ($p=2$) 
 and higher monomials for $p>2$, respectively. 

The anharmonic Hooke force (\ref{force}) implemented with
 the conservation law (\ref{1sti})
\begin{equation}\label{indat}
\sqrt{1- \dot{\mathbf{q}}^2}= \frac{T}{C}|q_{1} q_{2}...q_{p}|,
\end{equation}
restricts the character of the $p$-brane motion by 
\begin{equation}\label{restr}
0 \leq |\dot{\mathbf{q}}| \leq 1,  \ \ \  0 \leq \frac{T}{C}|q_{1} q_{2}...q_{p}|  \leq 1.
\end{equation}
The inequalities (\ref{restr}) imply that the velocity 
value $|\dot{\mathbf{q}}|$ grows when the $p$-brane (hyper) 
volume $\sim |q_{1} q_{2}...q_{p}|$ diminishes,  
and reaches the velocity of light ($|\dot{\mathbf{q}}|=1)$ 
while the (hyper)volume vanishes.
On the contrary, the  minimal velocity $\dot{\bf{q}}=0$ 
corresponds to the maximal (hyper)volume $\sim |q_{1} q_{2}...q_{p}|$ 
 equal to $C/T$. 
 
 For the case $T=0$, associated with the tensionless 
 p-branes  \cite{Z_0, BZ_0}, Eqs. (\ref{qeqv}) take the linear form
\begin{equation}\label{particl}
  \ddot{\mathbf{q}}=0,  \ \ \  |\dot{\mathbf{q}}| =1
\end{equation} 
similar to the equation of free massless particle in the 
effective space formed by the $p$-torus radii. 

For the tension $T$ different from zero the system (\ref{qeqv})
 is rather complicated, and its general solution is unknown. 
However, one can observe a case when equations (\ref{qeqv}) 
may be exactly solved.

\section{On integrability of $p$-brane equations}

Here we show the exact solvability of the nonlinear $p$-brane
 equations (\ref{qeqv})
for a degenerate case when all the components of $\mathbf{q}$ 
are equal: \mbox{$q_{1}= q_{2}=...=q_{p}\equiv q$}.
In this case the system (\ref{qeqv}) is reduced to the nonlinear
 differential equation 
\begin{equation}\label{eleq}
\ddot{q} + (\frac{T}{C})^2 q^{(2p-1)} = 0
\end{equation} 
which integration results in the first integral 
\begin{equation}\label{hyper} 
p\dot{q}^{2} + (\frac{T}{C})^{2}q^{2p} = 1.
\end{equation} 
After the  change of $q$ by \mbox{$y \equiv \Omega^{\frac{1}{p}}\sqrt{p}q$},
with $\Omega\equiv\frac{T}{C}p^{-\frac{p}{2}}$, 
Eq. (\ref{hyper}) takes the form
\begin{equation}\label{hypertr} 
(\frac{d y}{d\tilde{t}})^{2}= \frac{1}{2}(1-y^p)(1+y^p)
\end{equation} 
with the new time variable  $\tilde{t}\equiv \sqrt{2}\Omega^{\frac{1}{p}}t$. 

For the degenerate toroidal membranes ($p=2$) 
Eq. (\ref{hypertr}) coincides with the canonical equation defining 
the Jacobi elliptic cosine $cn(x;k)$
\begin{equation}\label{elcndef} 
(\frac{d y}{d x})^{2}= (1-y^2)(1- k^2 + k^2y^2),
 \end{equation} 
with the elliptic modulus $k=\frac{1}{\sqrt{2}}$.
 
 Thus, the general solution of (\ref{elcndef}) is 
 $$
 y(t)= cn(\sqrt{2\omega}t;\frac{1}{\sqrt{2}})
 $$ 
with $2\omega= T/C$.
  After using the relation $q\equiv y/{\sqrt{2\omega}}$ we obtain 
 the general solution for the coordinate $q(t)$ 
\begin{equation}\label{elcn} 
q(t)=
\sqrt{\frac{C}{T}}cn(\sqrt{\frac{T}{C}}(t+t_{0});\frac{1}{\sqrt{2}}).
\end{equation} 
This solution  is similar to the elliptic one earlier obtained in \cite{TZ, ZT} and 
describing the  $U(1)$ invariant membrane in the five-dimensional (i.e. $D=5$) Minkowski space.
If the initial velocity $\dot{q}(t_0)>0$,
the solution (\ref{elcn}) describes an expanding torus which reaches the maximal 
size $q_{max}=\sqrt{\frac{C}{T}}$ at
some moment $t$, and then contracts to a point after the finite time
$\mathbf{K}(1/\sqrt{2})\sqrt{\frac{C}{T}}$
(where $\mathbf{K}(1/\sqrt{2})=1.8451$) is the quarter period of elliptic
cosine).

An explicit equation of the surface $\Sigma_{2}(t)$ of the
 contracting torus (\ref{elcn}) is
 \begin{eqnarray}\label{surf}
 x_1^2+x_2^2+x_3^2+x_4^2&=&
\frac{4C}{T}cn\left(\sqrt{\frac {T}{C}}(t+t_0),\frac{1}{\sqrt{2}}\right)^2, \nonumber 
\\
 x_1x_4&=&x_2x_3.  
\end{eqnarray}
For the case $p>2$ integration of Eq. (\ref{hypertr}) results in the solution
\begin{equation}\label{solhyper} 
\tilde{t}=\pm\sqrt{2}\int\frac{dy}{\sqrt{1-y^{2p}}} + const
\end{equation} 
that contains hyperelliptic integral and defines implicit dependence of $q$ on the time.
Thus, the general solution of Eq. (\ref{hyper}) 
is expressed  in terms of hyperelliptic functions generalizing elliptic functions.  

The variable change $z=y^{2p}$ transforms the solution (\ref{solhyper}) into 
the integral
\begin{equation}\label{solhyp2} 
\tilde{t}-\tilde{t_0} 
= \pm\frac{1}{\sqrt{2}p}\int_{0}^{z^\frac{1}{2p}}dz z^{(\frac{1}{2p}- 1)}(1-z)^{-\frac{1}{2}}
\end{equation} 
similar to the integral discussed in \cite{LLm}.

The use of the representation (\ref{solhyp2}) allows to find the contraction 
time $\Delta\tilde{t}_{c}$ of the 
degenerate $p$-torus from its maximal size  $q_{max}=(\frac{C}{T})^\frac{1}{p}$ 
to $q_{min}=0$. 

This time turns out to be proportional to the well-known integral 
$$
\Delta\tilde{t}_{c}=\frac{1}{\sqrt{2}p}\int_{0}^{1}dz z^{(\frac{1}{2p}- 1)}(1-z)^{\frac{1}{2}-1}= \frac{1
}{\sqrt{2}p}B(\frac{1}{2p}, \frac{1}{2})
$$
which defines the Euler beta function $B(\frac{1}{2p}, \frac{1}{2})$.
 
Coming back to the original time  $t$ and taking into account that $C=2E$ 
we obtain  
\begin{equation}\label{period}
\Delta t_{c}\equiv\frac{1}{\sqrt{2}}\Omega^{-\frac{1}{p}}\Delta\tilde{t}_{c}
=\frac{1}{2\sqrt{p}}(\frac{2E}{T})^{\frac{1}{p}}
B(\frac{1}{2p}, \frac{1}{2}).
\end{equation} 
The representation (\ref{period}) gives the explicit dependence of  
the  contraction time 
in term sof of the $p$-brane dimension $p$ and its energy density $E$.

So, we conclude that  the case of the degenerate toroidal $p$-branes 
with coinciding radii is exactly solvable and connects the solutions 
of the $p$-brane equations with (hyper)elliptic functions.

\begin{itemize}
 
\item  
A special class of relativistic $p$-branes 
 embedded in the $D=(2p+1)$-dimensional Minkowski space is
 introduced for studying the elastic forces associated with branes.
The compact hypersurfaces 
of these  $p$-branes are chosen to be invariant 
under the transformations of the
 $O(2)\times O(2)\times\ldots \times O(2)$ subgroup 
of the  rotations of $2p$-dimensional Euclidean space.
\item 
The brane equations are found to be reduced to equations 
of the anharmonic elastic media subjected 
 to isotropic pressure dependent of time.
Their Hamiltonians including monomial potentials
of the degree $2p$ ($p=2,3,...,(D-1)/2$) and yielding nonlinear
  Hooke forces are constructed. 
 \item 
 The $p$-brane equations are proved to be exactly solvable if the brane shapes 
 are similar to $p$-tori with equal radii. 
The constructed  (hyper)elliptic solutions describe contracting $p$-tori. 
The exact formula for the contraction time of the toric  $p$-branes is derived.
In particular, these results give the new information on the nonlinear 
elastic potentials associated with five-branes ($p=5$) of mysterious 
M/string theory supposed to exist in the space-time with the exclusive
 dimension $D=11$. 
 
 In particular, these results give the new information on the nonlinear 
elastic potentials associated with five-branes ($p=5$) of mysterious 
M/string theory supposed to exist in the space-time with the exclusive
 dimension $D=11$. 
 Interestingly, a breakdown of the linear Hooke elasticity and its replacement 
 by a nonlinear anharmonic law, similar to the ones revealed by us,
 were earlier discovered in 2d and 3d smectics $\mathcal{A}$ (see e.g. \cite{GP}, \cite{GZ}).
 
 \end{itemize}
 
 \noindent{\bf Acknowledgments}

 I am grateful to J. Hoppe, V. Nesterenko, K. Stelle, M. Trzetrzelewski and 
 D. Uvarov for useful discussions and to Physics Department of Stockholm University,
Nordic Institute for Theoretical Physics Nordita for kind hospitality and financial support.
 
\begin{center}

\end{center}
\end{document}